\documentclass{aa}  
\usepackage{graphicx}
\usepackage{txfonts}
\usepackage{natbib}
\bibpunct{(}{)}{;}{a}{}{,} 

\newcommand{\pmj}{PM~J22299+3024}
\newcommand{\lp}{LP~119-10}

\usepackage[normalem]{ulem}

\begin{document}

   \title{Exploring the pulsational properties of two ZZ~Ceti stars}


    \author{Zs.~Bogn\'ar\inst{1,2,3}\fnmsep\thanks{\email{bognar@konkoly.hu}},
        Cs.~Kalup\inst{1,4},
        \and \'A.~S\'odor\inst{1,2}}
   
   \institute{
        Konkoly Observatory, E\"otv\"os Lor\'and Research Network (ELKH), Research Centre for Astronomy and Earth Sciences, Konkoly Thege Mikl\'os \'ut 15-17, H--1121, Budapest, Hungary
        \and
        MTA CSFK Lend\"ulet Near-Field Cosmology Research Group
        \and
        ELTE E\"otv\"os Lor\'and University, Institute of Phyiscs, P\'azm\'any P\'eter s\'et\'any 1/A, H-1171, Budapest, Hungary
        \and
        ELTE E\"otv\"os Lor\'and University, Department of Astronomy, P\'azm\'any P\'eter s\'et\'any 1/A, H-1171, Budapest, Hungary
        }
        
    \titlerunning{Exploring PM~J22299+3024 and LP~119-10}
        \authorrunning{Zs.~Bogn\'ar et al.}
        
    \date{}

 
  \abstract
   {We continued our ground-based observing project with the season-long observations of ZZ~Ceti stars at Konkoly Observatory. Our present targets are the newly discovered \pmj, and the already known \lp\ variables. \lp\ was also observed by the \textit{TESS} (\textit{Transiting Exoplanet Survey Satellite}) space telescope in 120-second cadence mode.}
   {Our main aims are to characterise the pulsation properties of the targets, and extract pulsations modes from the data for asteroseismic investigations.}
   {We performed standard Fourier analysis of the daily, weekly, and the whole data sets, together with test data of different combinations of weekly observations. We then performed asteroseismic fits utilising the observed and the calculated pulsation periods. For the calculations of model grids necessary for the fits, we applied the 2018 version of the White Dwarf Evolution Code. 
   }
   {We derived six possible pulsation modes for \pmj, and five plus two \textit{TESS} pulsation frequencies for \lp. Note that further pulsation frequencies may be present in the data sets, but we found their detection ambiguous, so we omitted them from the final frequency list. 
  Our asteroseismic fits of \pmj\ give $11\,400$\,K and $0.46\,M_{\sun}$ for the effective temperature and the stellar mass. The temperature is $\approx 800\,$K higher, while the mass of the model star is exactly the same as it was earlier derived by spectroscopy. Our model fits of \lp\ put the effective temperature in the range of $11\,800 - 11\,900\,$K, which is again higher than the spectroscopic $11\,290\,$K value, while our best model solutions give $M_* = 0.70\,M_{\sun}$ mass for this target, near to the spectroscopic value of $0.65\,M_{\sun}$, likewise in the case of \pmj. The seismic distances of our best-fitting model stars agree with the \textit{Gaia} astrometric distances of \pmj\ and \lp\ within the errors, validating our model results.
  }
   {}


   \maketitle
%

\section{Introduction}

About 97 per cent of the stars, including our Sun, will finally end their evolution as white dwarfs. Some of the white dwarf stars show low-amplitude, short-period light variations. These can be found at specific parts of the Hertzsprung--Russell diagram, and form three large groups: these are the GW~Vir, the V777~Her (DBV), and the ZZ~Ceti (DAV) variables. The hottest objects are the GW~Vir \mbox{(pre-)white} dwarfs with $\sim$80\,000--180\,000\,K effective temperatures and hydrogen deficient atmospheres, while the DBV and DAV stars are much cooler, with 22\,000--32\,000\,K and 10\,500--13\,000\,K effective temperatures, respectively, and their atmospheres are dominated by neutral helium (DBV) or hydrogen (DAV). For a summary on the characteristic of the different families of white dwarf pulsators, see the review of \citet{2019A&ARv..27....7C}.

The most populous group is that of ZZ~Ceti, as about 80 per cent of the known pulsating white dwarfs belong to this group. Besides these, new groups of pulsating white dwarf stars have been identified recently, such as the extremely low-mass DA pulsators (ELM-DAVs; \citealt{2012ApJ...750L..28H}), the extremely low-mass pulsating pre-white dwarf stars (pre-ELM WD variables; \citealt{2013Natur.498..463M}), and the so-called hot DAV stars \citep{2008MNRAS.389.1771K, 2013MNRAS.432.1632K, 2020MNRAS.497L..24R} located at $\sim 30\,000\,$K effective temperatures. Light variations were also detected in DQV variables with atmospheres rich in helium and carbon \citep{2008ApJ...678L..51M}. However, the observed variability of DQ objects could be explained by effects other than global pulsations, e.g. rapid rotation \citep{2016ApJ...817...27W}. ZZ~Ceti variables in detached white dwarf plus main-sequence (MS) binaries have also become known \citep{2015MNRAS.447..691P}. For comprehensive reviews on the characteristics of pulsating white dwarf stars, see the papers of \citet{2008ARA&A..46..157W}, \citet{2008PASP..120.1043F}, \citet{2010A&ARv..18..471A}, \citet{2019A&ARv..27....7C}, and \citet{2020FrASS...7...47C}. 

Compact stars, such as white dwarfs, are unique space laboratories. However, the only way we can study their internal structure is by investigating the excited waves propagating thorough their interiors, by means of asteroseismology. This makes the search for new pulsators among white dwarfs an important effort. This is why we have initiated a survey searching for new pulsating white dwarf targets for the \textit{TESS} (\textit{Transiting Exoplanet Survey Satellite}) space telescope \citep{2018MNRAS.478.2676B,2019AcA....69...55B}. One of our new discoveries was \pmj, a new pulsator candidate \citep{2019AcA....69...55B}.

Pulsation modes detected in such objects are low horizontal-degree ($\ell = 1$ and $2$), low-to-mid radial-order $g$-modes with periods ranging from a couples of minutes to about half an hour, and with amplitudes in the millimagnitude range. The periods of these modes are sensitive to the global stellar structure, the stellar rotation, the inner chemical stratification, and the dynamical processes operating in them, which highlights the great potential of asteroseismological investigations.

We note that different pulsational behaviour is observed at different parts of the ZZ~Ceti instability strip. While the hotter objects are more likely to show pulsation frequencies with stable amplitudes and phases, this changes as we investigate objects closer to the red edge (at lower effective temperatures) of the instability domain. At this part, short-term (days--weeks-long) amplitude and phase changes are more common, while we detect longer-period and larger-amplitude pulsations than in the hotter objects. Thanks to the \textit{Kepler} observations, the so-called outburst events were also exposed in such objects, which means recurring increases in the stellar flux (up to 15 per cent) in cool ZZ~Ceti stars (see e.g. \citealt{2017ASPC..509..303B}). This phenomenon might be in connection with the cessation of pulsations at the empirical red edge of the ZZ~Ceti instability strip \citep{2015ApJ...810L...5H}. 

The study of white dwarf stars contributes to the understanding of star formation and evolution, and in addition, by investigating their interiors, we can use them as cosmic laboratories to study the behaviour of material under extreme pressure and temperature conditions, and measure the age of their parent stellar population. We know about 260 ZZ~Ceti stars \citep{2020FrASS...7...47C} currently, nonetheless, only a limited number of pulsation modes are known for most of them, usually the results of the short discovery runs. This is mainly because of the faintness of these objects, and because of the limited access of ground-based telescopes large enough for follow-up observations. However, we need more pulsation modes for asteroseismology for sufficient constraints on the physical parameters of the stars. Fortunately, there are several ways to collect more information on pulsating white dwarf stars. 

International campaigns, such as the Whole Earth Telescope (WET; \citealt{1990ApJ...361..309N}) proved already that we can extract sufficient number of pulsation frequencies by such observations for performing asteroseismic modelling. Another way is to utilise space-based time-series photometry of white dwarf variables. These space-based observations gave boost to the investigations of such objects. During the nominal \textit{Kepler} mission and its \textit{K2} extension, 81 ZZ~Ceti stars were observed, and the analyses of 32 of them have been published so far, see e.g. \citet{2017ApJS..232...23H, 2017ApJ...841L...2H, 2017ApJ...851...24B}, and \citet{2020FrASS...7...47C}. Published results have already demonstrated the value of the \textit{TESS} data, too, focusing on a DBV star \citep{2019A&A...632A..42B}, several ZZ~Ceti stars \citep{2020A&A...638A..82B, 2020A&A...633A..20A}, and GW~Vir variables \citep{2021A&A...645A.117C}.

We follow a third way, and perform long-term single-site ground-based observations of selected targets not observed extensively before, such as in the case of \lp, presented in this paper. Considering LP~119-10, only the result of the discovery run presenting one period has been published so far. With this publication on \pmj\ and \lp, we continue our efforts to introduce the results of our long-term ground-based observations on pulsating white dwarf stars in a series of papers, see e.g. \citet{2009MNRAS.399.1954B}, \citet{2013MNRAS.432..598P}, \citet{2014A&A...570A.116B}, \citet{2016MNRAS.461.4059B}, and \citet{2019MNRAS.482.4018B}. 

\section{Observations and data reduction}

We performed the observations with the 1-m Ritchey--Chr\'etien--Coud\'e telescope located at the Piszk\'estet\H o mountain station of Konkoly Observatory, Hungary. We obtained data with an FLI Proline 16803 CCD camera in white light. The exposure times were selected to be 45\,s and 30\,s in most cases for PM~J22299+3024 (fainter target) and LP~119-10, respectively. We applied longer exposures, up to 60\,s, in the case of unfavourable weather conditions. The read-out time was $\sim$3\,s.

We reduced the raw data frames the standard way utilising \textsc{iraf}\footnote{\textsc{iraf} is distributed by the National Optical Astronomy Observatories, which are operated by the Association of Universities for Research in Astronomy, Inc., under cooperative agreement with the National Science Foundation.} tasks: we performed bias, dark and flat corrections before the aperture photometry of field stars. We fitted low-order (second- or third-order) polynomials to the resulting light curves, correcting for long-period instrumental and atmospheric trends. This procedure did not affect the known frequency domain of pulsating ZZ~Ceti stars, however, made the detection of any possible long-period light variations, e.g. outburst events, difficult or even impossible. Finally, we converted the observational times of every data point to barycentric Julian dates in barycentric dynamical time (BJD$_\mathrm{{TDB}}$) using the applet of \citet{2010PASP..122..935E}\footnote{	http://astroutils.astronomy.ohio-state.edu/time/utc2bjd.html}.

Tables~\ref{tabl:logpmj} and \ref{tabl:loglp} show the journals of observations of PM~J22299+3024 and LP~119-10, respectively. We collected data on 14 and 6 nights in the 2018 and 2019 observing seasons on PM~J22299+3024, respectively, covering 97 and almost 40 hours with our measurements, while we observed LP~119-10 on 15 nights in one season, which resulted in the collection of 86 hours of photometric data on this target.

\begin{table}
\centering
\caption{Journal of observations of PM~J22299+3024. `Exp' is the integration time used, \textit{N} is the number of data points, and $\delta T$ is the length of the data sets including gaps. Weekly observations are denoted by `a,b,c,d,e,f,g' letters in parentheses.}
\label{tabl:logpmj}
\tiny
\begin{tabular}{lrccrr}
\hline
\hline
Run & UT Date & Start time & Exp. & \textit{N} & $\delta T$ \\
 &  & (BJD-2\,450\,000) & (s) &  & (h) \\
\hline
01(a) & 2018 Jul 20 & 8320.346 & 30 & 546 & 5.64 \\
02(b) & 2018 Sep 07 & 8369.268 & 30 & 779 & 8.77 \\
03(b) & 2018 Sep 10 & 8372.267 & 45 & 580 & 8.51 \\
04(b) & 2018 Sep 11 & 8373.280 & 45 & 596 & 8.41 \\
05(b) & 2018 Sep 12 & 8374.264 & 45 & 652 & 8.81 \\
06(c) & 2018 Sep 20 & 8382.248 & 45 & 600 & 8.35 \\
07(c) & 2018 Sep 21 & 8383.244 & 45 & 307 & 4.19 \\
08(d) & 2018 Oct 11 & 8403.221 & 45 & 550 & 7.49 \\
09(d) & 2018 Oct 12 & 8404.229 & 45 & 526 & 7.34 \\
10(d) & 2018 Oct 13 & 8405.238 & 45 & 532 & 7.12 \\
11(d) & 2018 Oct 14 & 8406.220 & 45 & 546 & 7.57 \\
12(e) & 2018 Nov 04 & 8427.188 & 40 & 475 & 6.37 \\
13(e) & 2018 Nov 06 & 8429.191 & 40 & 351 & 4.24 \\
14(e) & 2018 Nov 07 & 8430.182 & 40 & 360 & 4.49 \\
\multicolumn{2}{l}{Total:} & & \multicolumn{2}{r}{7400} & 97.30\\
\\
15(f) & 2019 Sep 21 & 8748.258 & 30 & 913 & 8.27 \\
16(f) & 2019 Sep 22 & 8749.260 & 60 & 322 & 6.24 \\
17(g) & 2019 Oct 24 & 8781.213 & 30 & 648 & 5.92 \\
18(g) & 2019 Oct 25 & 8782.217 & 30 & 742 & 6.72 \\
19(g) & 2019 Oct 26 & 8783.213 & 30 & 769 & 7.00 \\
20(g) & 2019 Oct 27 & 8784.215 & 30 & 625 & 5.65 \\
\multicolumn{2}{l}{Total:} & & \multicolumn{2}{r}{4019} & 39.80\\
\hline
\end{tabular}
\end{table}

\begin{table}
\centering
\caption{Journal of observations of LP~119-10. `Exp' is the integration time used, \textit{N} is the number of data points, and $\delta T$ is the length of the data sets including gaps. Weekly observations are denoted by `a,b,c,d,e,f,g' letters in parentheses.}
\label{tabl:loglp}
\tiny
\begin{tabular}{lrccrr}
\hline
\hline
Run & UT Date & Start time & Exp. & \textit{N} & $\delta T$ \\
 &  & (BJD-2\,450\,000) & (s) &  & (h) \\
\hline
01(a) & 2018 Oct 15 & 8407.388 & 30 & 696 & 6.74 \\
02(b) & 2018 Nov 03 & 8426.406 & 20 & 873 & 5.64 \\
03(b) & 2018 Nov 05 & 8428.377 & 30 & 657 & 7.35 \\
04(b) & 2018 Nov 06 & 8429.375 & 30 & 706 & 6.58 \\
05(b) & 2018 Nov 07 & 8430.374 & 30 & 804 & 7.48 \\
06(c) & 2018 Nov 30 & 8453.267 & 40 & 357 & 4.38 \\
07(c) & 2018 Dec 05 & 8458.250 & 30 & 1051 & 10.34 \\
08(d) & 2019 Jan 03 & 8487.235 & 30 & 376 & 3.68 \\
09(d) & 2019 Jan 07 & 8490.511 & 60 & 217 & 3.87 \\
10(d) & 2019 Jan 07 & 8491.216 & 30 & 318 & 3.03 \\
11(e) & 2019 Feb 07 & 8522.277 & 30 & 587 & 5.61 \\
12(e) & 2019 Feb 11 & 8526.346 & 30 & 505 & 4.72 \\
13(e) & 2019 Feb 12 & 8527.214 & 30 & 846 & 7.71 \\
14(f) & 2019 Mar 12 & 8555.242 & 30 & 649 & 5.87 \\
15(g) & 2019 Apr 06 & 8580.269 & 30 & 374 & 3.47 \\
\multicolumn{2}{l}{Total:} & & \multicolumn{2}{r}{9016} & 86.47\\
\hline
\end{tabular}
\end{table}

\begin{figure*}
\centering
\includegraphics[width=0.7\textwidth, angle=270]{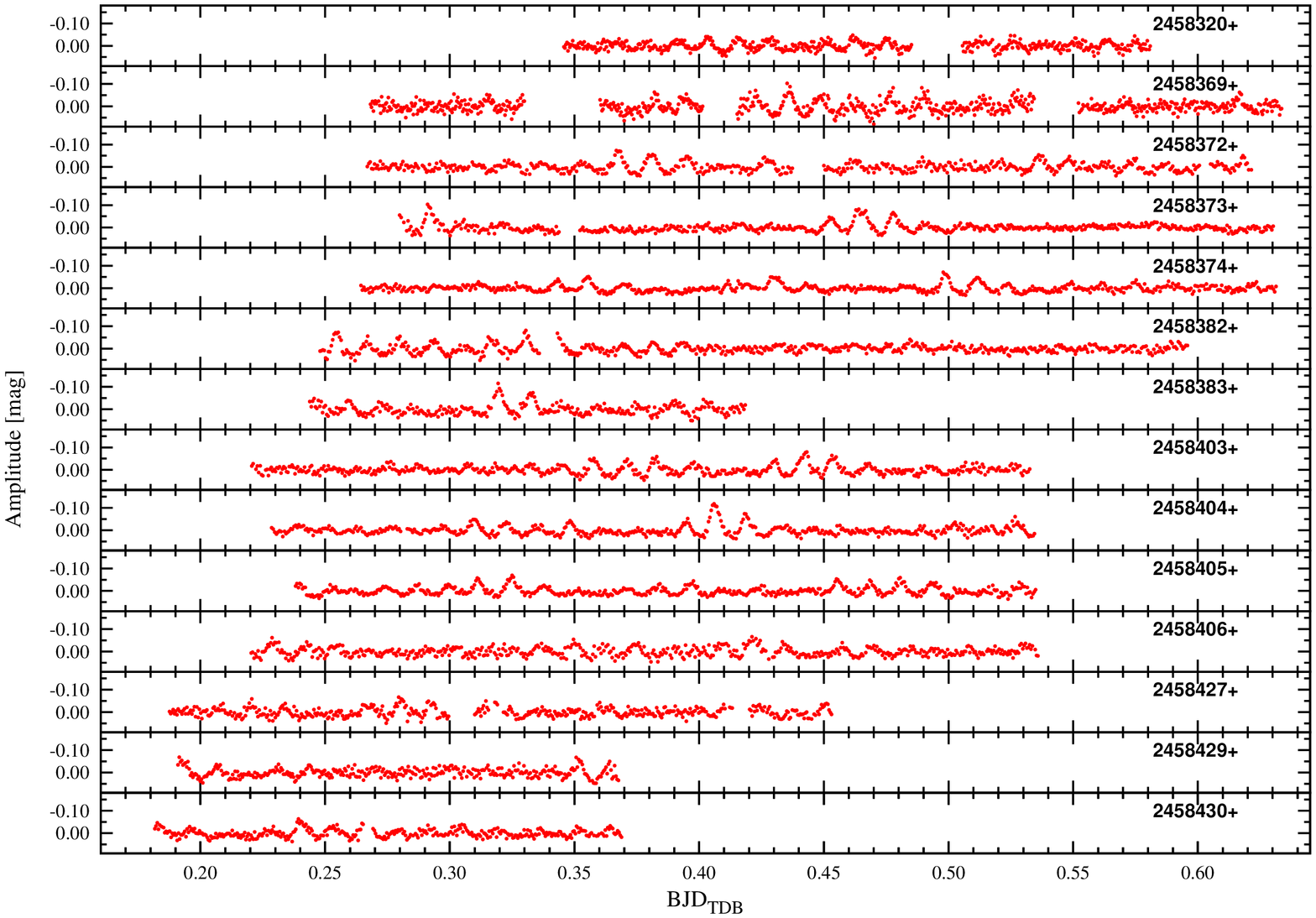}
\caption{Normalised differential light curves of PM~J22299+3024 obtained in the 2018 observing season.}{\label{fig:pmjbjd}}
\end{figure*}

\begin{figure*}
\centering
\includegraphics[width=1.0\textwidth, angle=0]{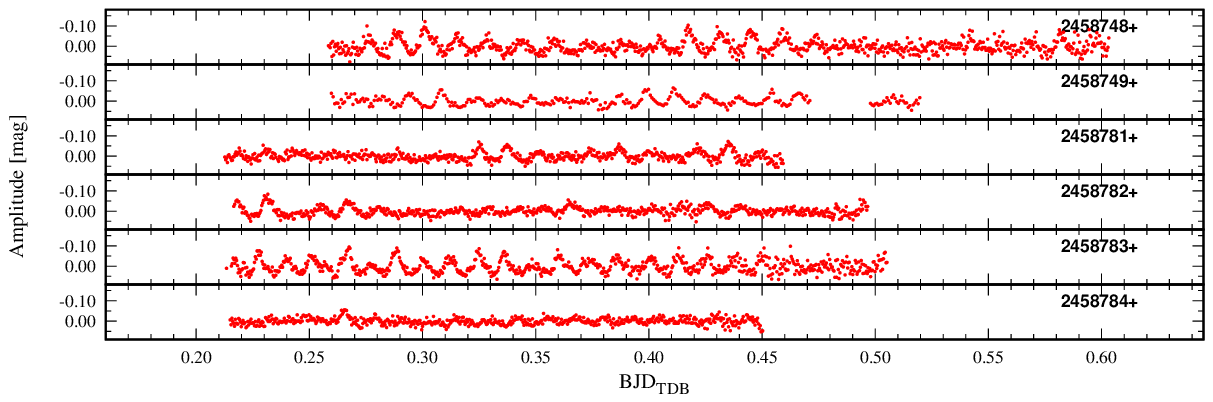}
\caption{Normalised differential light curves of PM~J22299+3024 obtained in the 2019 observing season.}{\label{fig:pmjbjd2}}
\end{figure*}

\begin{figure*}
\centering
\includegraphics[width=1.0\textwidth, angle=0]{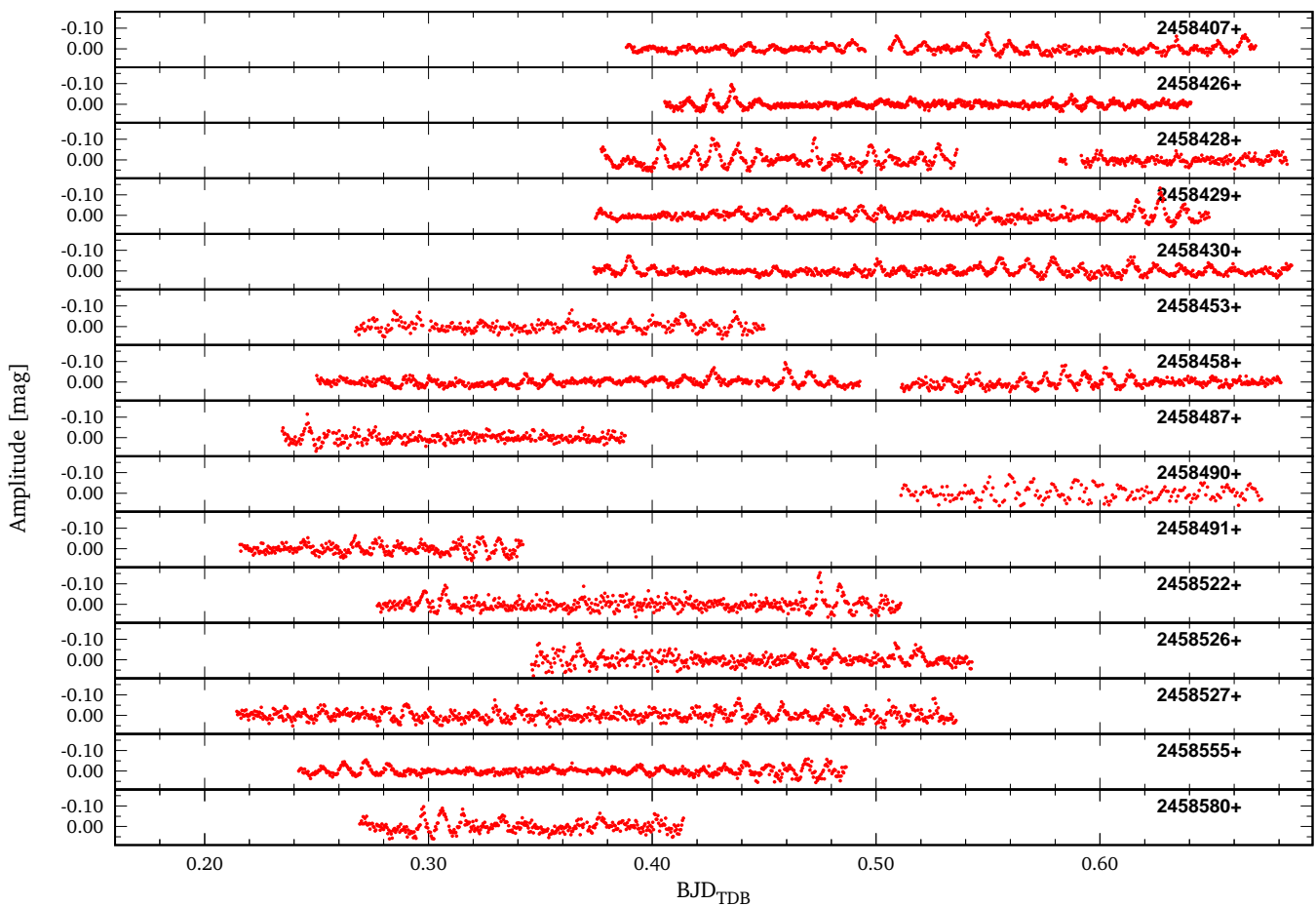}
\caption{Normalised differential light curves of LP~119-10.}{\label{fig:lpbjd}}
\end{figure*}

Figures~\ref{fig:pmjbjd} and \ref{fig:pmjbjd2} show the normalised differential light curves of PM~J22299+3024, respectively, while the plot of Fig.~\ref{fig:lpbjd} represents the ground-based light curves of LP~119-10.

\section{Light curve analysis}

We performed standard Fourier analysis on the data sets with the photometry modules of the Frequency Analysis and Mode Identification for Asteroseismology (\textsc{famias}) software package \citep{2008CoAst.155...17Z}. We accepted a frequency peak as significant if its amplitude reached the five signal-to-noise ratio (S/N), where the noise level was calculated by the average Fourier amplitude in a $\sim 1700\,\mu$Hz radius vicinity ($150\,$d$^{-1}$) of the peak in question (see e.g. in \citealt{2019MNRAS.482.4018B}). That is, we used higher significance level than the usual 4~S/N; we accepted the highest-amplitude peaks as possible pulsational frequencies during the pre-whitening process of these targets showing complex pulsational behaviour, with several closely spaced peaks in their Fourier transforms (FTs).

\subsection{PM~J22299+3024}

PM~J22299+3024 ($G=16.21$\,mag, $\alpha_{2000}=22^{\mathrm h}29^{\mathrm m}58^{\mathrm s}$, $\delta_{2000}=+30^{\mathrm d}24^{\mathrm m}10^{\mathrm s}$) was found to be a variable candidate by our research group in 2018 July \citep{2019AcA....69...55B}. We performed survey observations to find new bright white dwarf pulsators for the \textit{TESS} \citep{2015JATIS...1a4003R} all-sky survey space mission. At that time we considered it a variable candidate as only one night of observations was available on this target. However, the subsequent observations presented in this paper confirmed that PM~J22299+3024 is indeed a new, bright ZZ~Ceti star, situated close to the red edge of the instability strip according to spectroscopy \citep{2015ApJS..219...19L}.

First, we performed the Fourier analysis of the daily and weekly data sets, and finally, we analysed the complete 2018 and 2019 data sets, respectively. We also analysed data sets constructed by various combinations of different consecutive weekly data, testing our frequency solutions on data sets with different spectral windows. These six test data sets were consist of the data of weeks (a+b+c), (b+c+d), (c+d+e), (a+b+c+d), (b+c+d+e), and (f+g), cf. Table~\ref{tabl:logpmj}. 

Our set of accepted frequencies is based on the analysis of the combined weekly data subsets. These are listed in Table~\ref{tabl:pmjfreq2}. We accepted frequencies as real pulsation components that was found in at least three subsets.
We identified six pulsation frequencies in the $\sim750 - 960\,\mu$Hz frequency range. We did not find any combination frequencies.

\begin{table*}
\centering
\caption{The appearance of the accepted pulsation frequencies of PM~J22299+3024 in different combined weekly data subsets.}
\label{tabl:pmjfreq2}
\begin{tabular}{lrrrrrr}
\hline
\hline
 & \multicolumn{6}{c}{frequency [$\mu$Hz]} \\
\hline
week(a+b+c)	  &	749.3 &	837.0 &	855.2 & 885.0 &	922.0 &	959.5 \\
week(b+c+d)	  &	750.4 &	840.3 &	--    &	885.0 &	922.1 &	959.4 \\
week(c+d+e)	  &	--    &	839.2 &	853.1 &	884.4 &	921.0 &	960.0 \\
week(a+b+c+d) &	750.4 &	839.6 &	852.6 & 885.0 &	922.0 &	959.8 \\
week(b+c+d+e) &	749.3 &	--    & 852.6 &	885.0 &	922.0 &	960.5 \\
week(f+g)	  &	--    &	--    &	--    &	883.8 &	920.1 &	949.7 \\
& & & & & & \\
average       & 749.9 & 839.0 & 853.4 & 884.6 & 921.5 & 958.2 \\
\hline
\end{tabular}
\end{table*}

\begin{figure}
\centering
\includegraphics[width=0.47\textwidth]{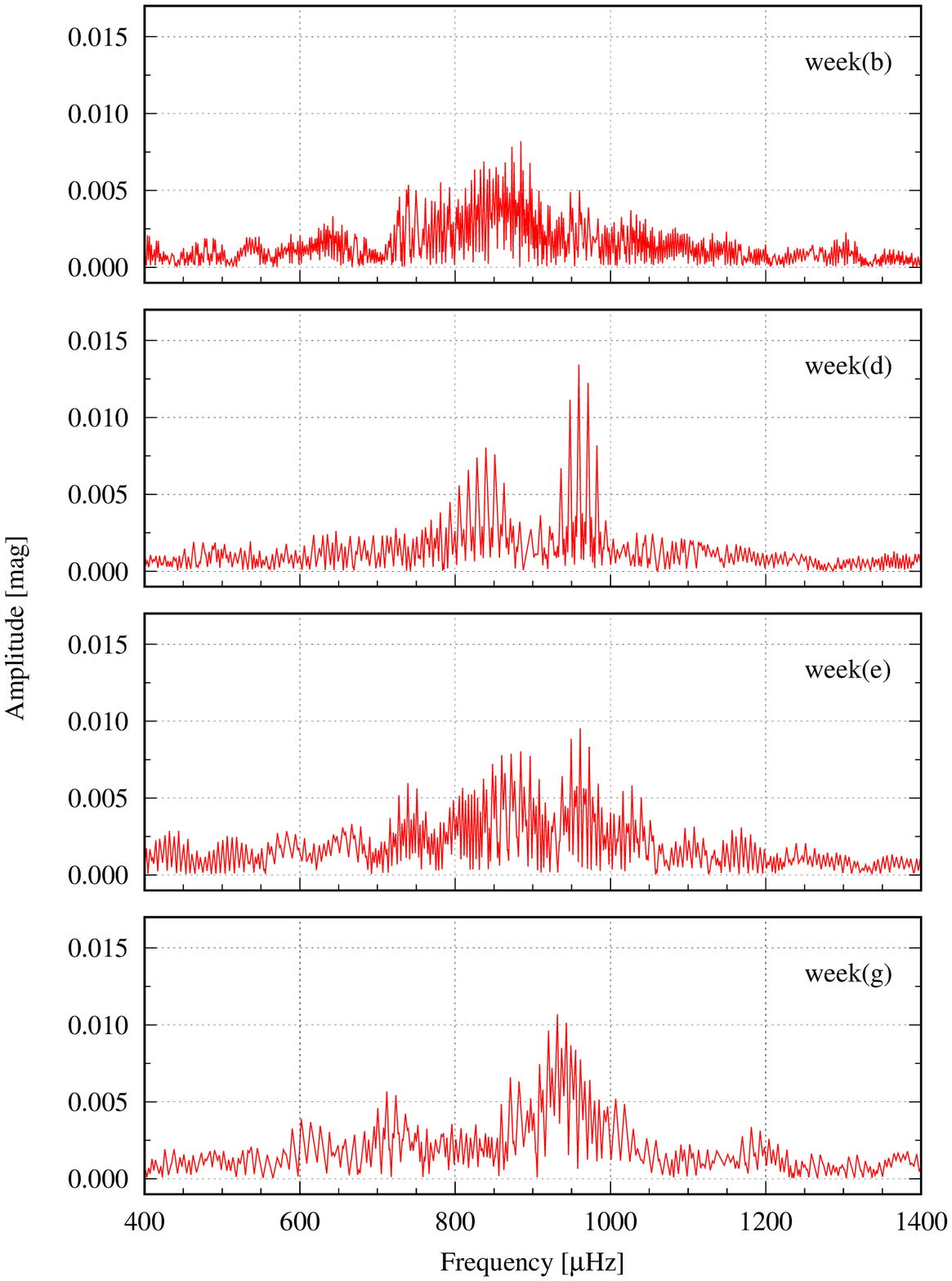}
\caption{PM~J22299+3024: Fourier transform of the weekly data sets with three or more nights of observations.}{\label{fig:weeklyftpmj}}
\end{figure}

Note that, besides the frequencies presented in Table~\ref{tabl:pmjfreq2}, further frequencies can also be identified in our data sets. There may be additional frequencies at around $790-800$, $820$, $860$, $900$, and $970\,\mu$Hz. However, the identification of these components and the establishment of their frequency values was ambiguous, therefore, we omitted them from the set of accepted pulsation frequencies listed in Table~\ref{tabl:pmjfreq2}. The ambiguities of the omitted components have possibly two main sources: $1\,$d$^{-1}$ aliasing, and short-term amplitude or phase variations. Figure~\ref{fig:weeklyftpmj} shows that variations in the amplitudes of the pulsation components occurred indeed from one observing week to another.

\begin{figure*}
\centering
\includegraphics[width=1.0\textwidth]{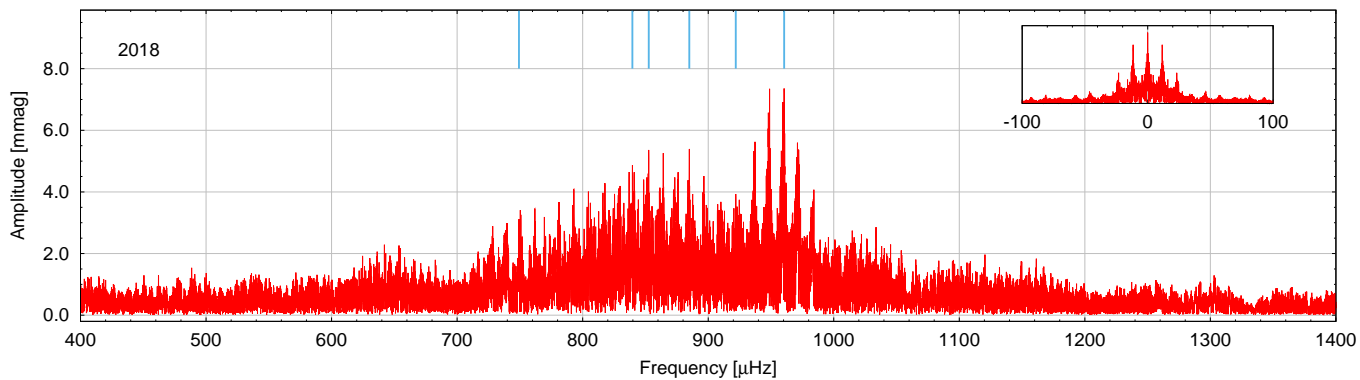}
\caption{PM~J22299+3024: Fourier amplitude spectrum of the complete 2018 data set. We marked the accepted frequencies listed in Table~\ref{tabl:pmjfreq} with blue lines. The window function is shown in the inset.}{\label{fig:wholeftpmj}}
\end{figure*}

One of the main goals of the frequency analysis was to provide periods for the asteroseismic models. The set of accepted frequency components is based on the findings presented in Table~\ref{tabl:pmjfreq2}. We refined the frequencies and amplitudes by fitting the complete 2018 data set by the six accepted components.
We utilised the obtained periods as input for the asteroseismic modelling described in Sect.~\ref{sect:asteropmj}. We present the Fourier amplitude spectrum of the complete 2018 data set in Fig.~\ref{fig:wholeftpmj}.    

\begin{table}
\centering
\caption{PM~J22299+3024: frequencies, periods and amplitudes of the six accepted pulsation components based on the 2018 observations. The components are listed in decreasing order of amplitude.
}
\label{tabl:pmjfreq}
\begin{tabular}{lrrrr}
\hline
\hline
 & \multicolumn{1}{c}{\textit{f}} & \multicolumn{1}{c}{\textit{P}} & Ampl. \\
 & \multicolumn{1}{c}{[$\mu$Hz]} & \multicolumn{1}{c}{[s]} & [mmag] \\
\hline
$f_1$ & 960.48 & 1041.14 & 7.3 \\
$f_2$ & 852.64 & 1172.83 & 5.6 \\
$f_3$ & 884.95 & 1130.00 & 5.3 \\
$f_4$ & 839.61 & 1191.03 & 5.1 \\
$f_5$ & 921.96 & 1084.65 & 3.7 \\
$f_6$ & 749.28 & 1334.61 & 2.9 \\
\hline
\end{tabular}
\end{table}

Space-based observations would definitely help to verify our Fourier solution and to resolve the frequency ambiguities. PM~J22299+3024 was on the list of proposed objects for \textit{TESS} measurements, and observations were predicted to be performed between 2019 September 11 and October 7 (cycle 2, sector 16) on this object. However, because of the unexpected field shifts of \textit{TESS}, the telescope did not observe PM~J22299+3024. This was one of the reasons why we decided to collect more data on this target from the ground in 2019.

\subsection{LP~119-10}

\lp\ ($G=15.26$\,mag, $\alpha_{2000}=05^{\mathrm h}02^{\mathrm m}34^{\mathrm s}$, $\delta_{2000}=+54^{\mathrm d}01^{\mathrm m}09^{\mathrm s}$) was found to be a variable DA-type white dwarf star by \citet{2015ASPC..493..237G}. They published one pulsation period for this object at $873.6$\,s with an amplitude of $1.27\%$. We observed the star on 15 nights in the 2018/2019 observing season.

\begin{figure*}
\centering
\includegraphics[width=1.0\textwidth]{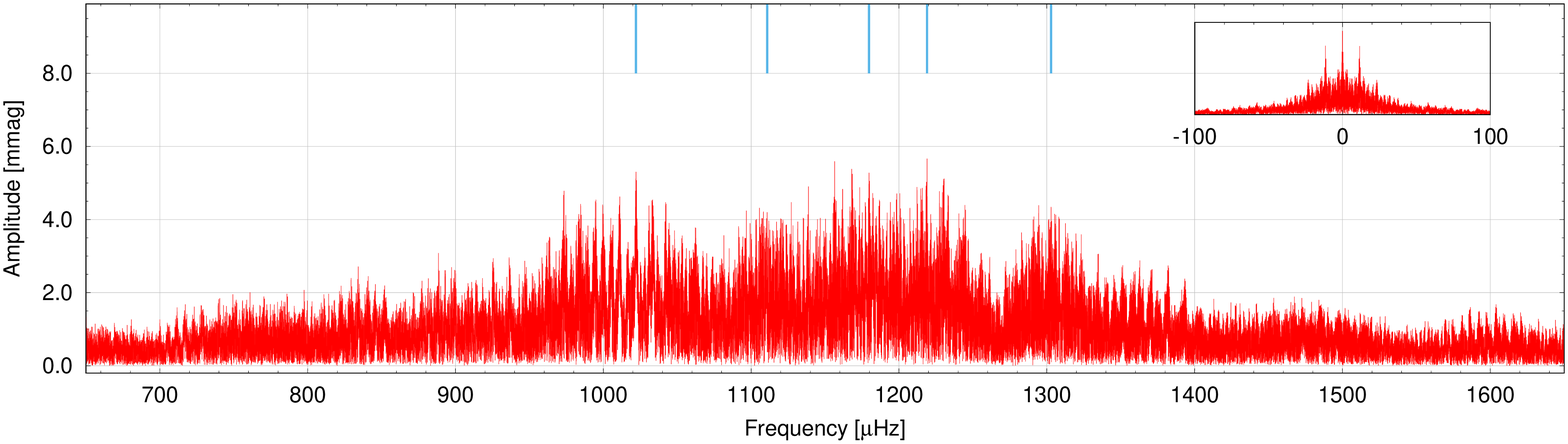}
\caption{\lp: Fourier amplitude spectrum of the complete data set. We mark the accepted frequencies listed in Table~\ref{tabl:lpfreq2} with blue lines. The window function is shown in the inset.}
{\label{fig:wholeftlp}}
\end{figure*}

Similarly to \pmj, we performed Fourier analysis not only on the daily, weekly, and the complete data sets, but also on different combinations of the weekly data. We constructed nine such data subsets combining weekly data sets of (a+b+c), (b+c+d), (c+d+e), (d+e+f), (e+f+g), (a+b+c+d), (b+c+d+e), (c+d+e+f), and (d+e+f+g). Table~\ref{tabl:lpfreq} lists the accepted components with peaks close in frequencies in at least four data subsets. As Table~\ref{tabl:lpfreq} shows, we identified five possible pulsation frequencies in the frequency range of $1020 - 1310\,\mu$Hz. Furthermore, similarly to \pmj, other possible pulsation frequencies are suspected at around $970 - 995$, $1040$, $1155$, $1195$, and $1225-1245\,\mu$Hz. Further observations may lead to a more solid identification of these components. We refined the parameter of the five accepted pulsation components by fitting the complete data set. The results are listed in Table~\ref{tabl:lpfreq2}, while we present the Fourier amplitude spectrum of the complete \lp\ data set in Fig.~\ref{fig:wholeftlp}. 

\begin{table*}
\centering
\caption{Values of the accepted pulsation frequencies of \lp\ derived by the different combined weekly data subsets.}
\label{tabl:lpfreq}
\begin{tabular}{lrrrrr}
\hline
\hline
 & \multicolumn{5}{c}{frequency [$\mu$Hz]} \\
\hline
week(a+b+c)	  & 1022.5 & 1110.3 & --	 & 1214.7 &	1301.9 \\
week(b+c+d)	  & 1022.5 & --     & 1180.5 & 1215.1 &	1301.4 \\
week(c+d+e)	  &	--     & 1110.1	& 1180.0 & --	  &	1305.9 \\
week(d+e+f)	  &	--	   & --     & 1180.0 & 1222.2 & 1308.5 \\
week(e+f+g)	  &	--     & 1114.7	& 1182.2 & --     & 1308.5 \\
week(a+b+c+d) &	1022.4 & 1110.8	& --	 & 1214.7 &	1303.8 \\
week(b+c+d+e) &	1022.1 & 1110.8	& 1179.7 & 1218.7 & 1306.2 \\
week(c+d+e+f) &	--	   & 1110.5 & 1180.1 & 1219.0 &	1308.5 \\
week(d+e+f+g) &	--	   & 1110.1	& 1180.0 & 1222.1 &	1308.5 \\
 & & & & & \\							
average	      & 1022.4 & 1111.1	& 1180.4 & 1218.0 & 1305.9 \\
\hline
\end{tabular}
\end{table*}

\begin{table}
\centering
\caption{\lp: list of the set of accepted frequencies based on the whole 2018--2019 data set. The frequencies are listed in decreasing order of amplitude.
}
\label{tabl:lpfreq2}
\begin{tabular}{lrrrr}
\hline
\hline
 & \multicolumn{1}{c}{\textit{f}} & \multicolumn{1}{c}{\textit{P}} & Ampl. \\
 & \multicolumn{1}{c}{[$\mu$Hz]} & \multicolumn{1}{c}{[s]} & [mmag] \\
\hline
$f_1$ & 1218.99 & 820.35 & 6.2 \\
$f_2$ & 1179.72 & 847.66 & 5.7 \\
$f_3$ & 1022.06 & 978.42 & 5.4 \\
$f_4$ & 1302.90 & 767.52 & 4.4 \\
$f_5$ & 1110.84 & 900.22 & 3.7 \\
\hline
\end{tabular}
\end{table}

\subsubsection{\textit{TESS} observations}

\begin{figure}
\centering
\includegraphics[width=0.47\textwidth]{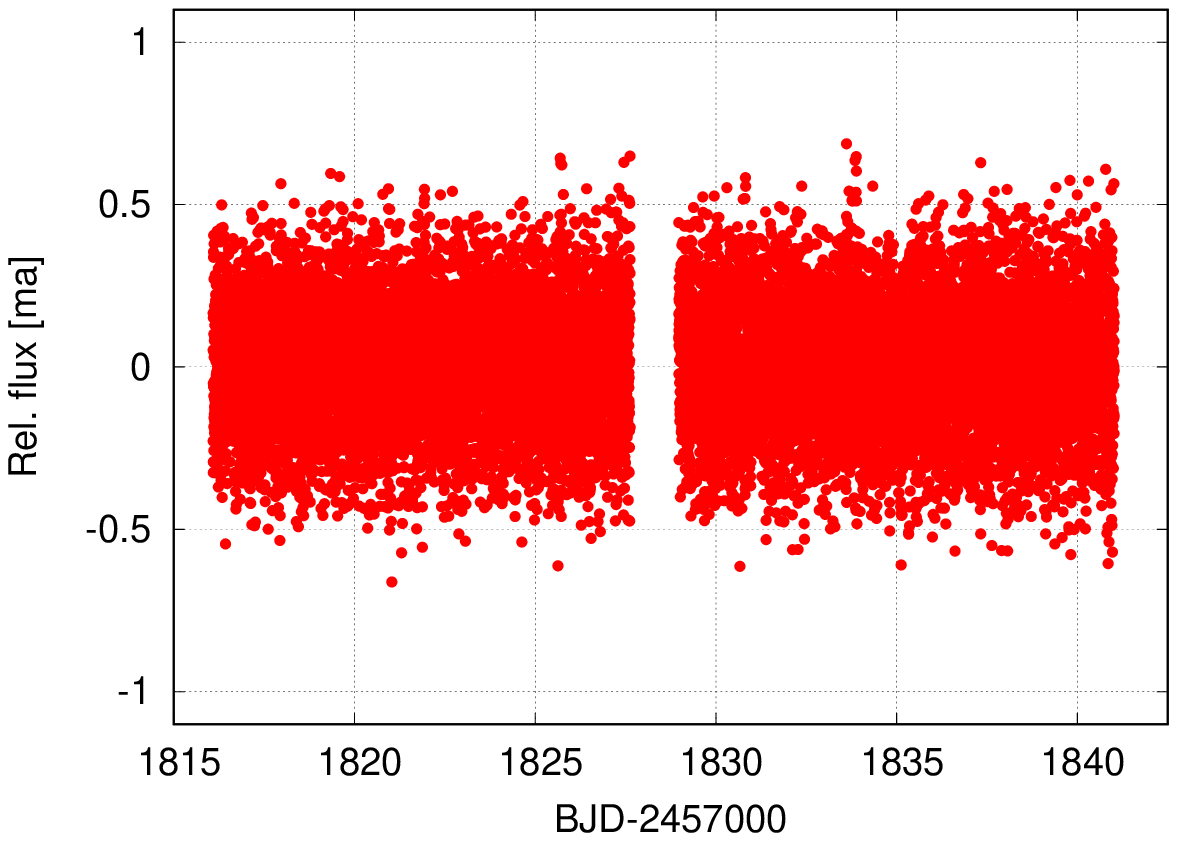}
\caption{\textit{TESS} light curve of \lp.}{\label{fig:tesslc}}
\end{figure}

\textit{TESS} observed \lp\ for $24.9$ days in sector 19 with the 120-second short-cadence mode. We downloaded the light curves from the \textit{Mikulski Archive for Space Telescopes} (MAST), and extracted the PDCSAP fluxes provided by the Pre-search Data Conditioning Pipeline \citep{2016SPIE.9913E..3EJ}.  We omitted the obvious outliers. The resulting light curve consists of 16\,494 data points (with a gap), as it can be seen on Fig~\ref{fig:tesslc}.

\begin{figure}
\centering
\includegraphics[width=0.47\textwidth]{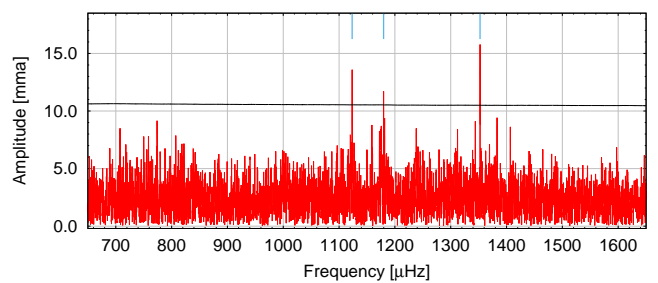}
\caption{\lp: Fourier transform of the \textit{TESS} light curve. We marked the frequencies listed in Table~\ref{tabl:lpfreqtess} with blue lines. The black line denotes the 4~S/N significance level.}{\label{fig:ftTESS}}
\end{figure}

\begin{figure}
\centering
\includegraphics[width=0.47\textwidth]{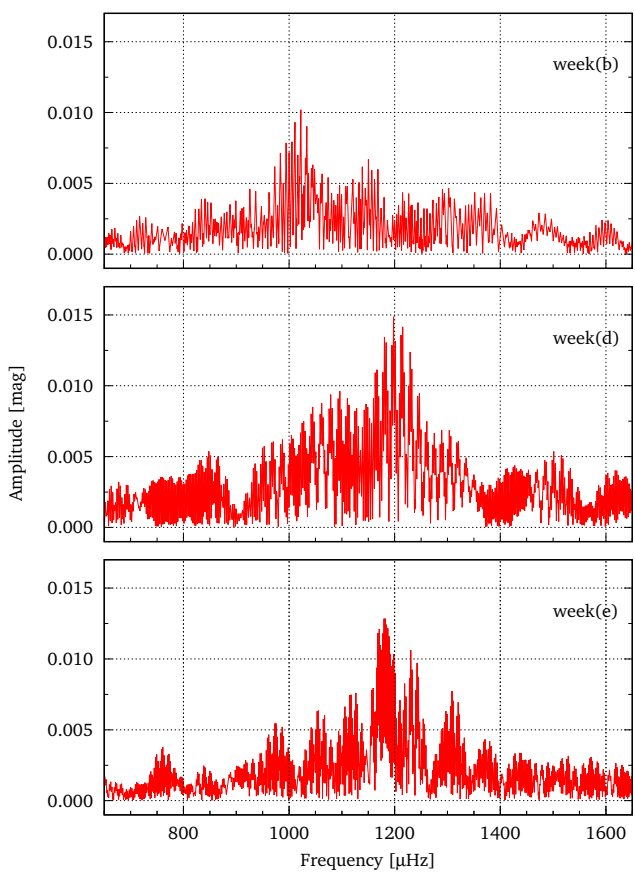}
\caption{\lp: Fourier transform of the three weekly data sets consist of more than two nightly runs.}{\label{fig:weeklyftlp}}
\end{figure}

The Fourier analysis of the \textit{TESS} data revealed three significant frequencies above the 4~S/N limit, listed in Table~\ref{tabl:lpfreqtess}. Comparing the frequency contents of the ground-based and space-based observations, we can find one common frequency ($f_2 = f_{3,TESS}$). The other two frequencies are new detections, including the dominant \textit{TESS} frequency. This suggests amplitude variations on time scales of months, supported by the different Fourier transforms of the weekly data sets plotted in Fig.~\ref{fig:weeklyftlp}. Figure~\ref{fig:ftTESS} shows the Fourier transform of the \textit{TESS} data set.   

\begin{table}
\centering
\caption{\lp: set of of frequencies derived by the \textit{TESS} data set. The frequencies are listed in decreasing order of amplitude.
}
\label{tabl:lpfreqtess}
\begin{tabular}{lrrrr}
\hline
\hline
 & \multicolumn{1}{c}{\textit{f}} & \multicolumn{1}{c}{\textit{P}} & Ampl. \\
 & \multicolumn{1}{c}{[$\mu$Hz]} & \multicolumn{1}{c}{[s]} & [mma] \\
\hline
$f_{1,TESS}$ & 1352.58 & 739.33 & 15.8 \\
$f_{2,TESS}$ & 1123.42 & 890.14 & 13.6 \\
$f_{3,TESS}$ & 1179.79 & 847.66 & 11.8 \\
\hline
\end{tabular}
\end{table}

\section{Asteroseismology}

We built model grids for the asteroseismic investigations of both PM~J22299+3024 and LP~119-10, utilising the White Dwarf Evolution Code (\textsc{wdec}) version presented in 2018 \citep{2018AJ....155..187B}. This updated version of the \textsc{wdec} uses \textsc{mesa} (Modules for Experiments In Stellar Astrophysics; \citealt{2011ApJS..192....3P}, version r8118) equation of states and opacity routines.

The starting model is a hot ($\sim100\,000$\,K) polytrope, which is evolved down to the requested temperature. The model we finally obtain is a thermally relaxed solution to the stellar structure equations. The convection is treated within the mixing length theory \citep{1971A&A....12...21B}. We chose to use the $\alpha$ parametrization, according to the results of \citet{2015ApJ...799..142T}. 

We compute the set of possible $\ell=1$ and $2$ eigenmodes for each model according to the adiabatic equations of non-radial stellar oscillations \citep{1989nos..book.....U}. The goodness of the fit between the observed ($P_i^{\mathrm{obs}}$) and calculated ($P_i^{\mathrm{calc}}$) periods is characterised by the root mean square ($\sigma_\mathrm{{rms}}$) value calculated for every model with the \textsc{fitper} program of \citet{2007PhDT........13K}:

\begin{equation}
\sigma_\mathrm{{rms}} = \sqrt{\frac{\sum_{i=1}^{N} (P_i^{\mathrm{calc}} - P_i^{\mathrm{obs}})^2}{N}}
\label{equ1}
\end{equation}

\noindent where \textit{N} is the number of observed periods.

\subsection{The coarse (master) model grid}

At first, we built a coarse (master) model grid, covering a wide parameter space in effective temperature and stellar mass. For this, we varied six input parameters of the \textsc{wdec}: $T_{\mathrm{eff}}$, $M_*$, $M_\mathrm{{env}}$ (the mass of the envelope, determined by the location of the base of the mixed helium and carbon layer), $M_\mathrm{H}$, $X_\mathrm{{He}}$ (the helium abundance in the C/He/H region), and $X_\mathrm{O}$ (the central oxygen abundance). The second column of Table~\ref{tabl:grid} shows the parameter space we covered with the master grid, and the step sizes applied.

\begin{table*}
\centering
\caption{The parameter spaces covered by the master grid and the refined grids. The step sizes are in parentheses.}
\label{tabl:grid}
\begin{tabular}{lrrr}
\hline
\hline
& \multicolumn{1}{c}{master grid} & \multicolumn{1}{c}{refined grid -- \pmj} & \multicolumn{1}{c}{refined grid -- \lp} \\
\hline
$T_{\mathrm{eff}}$ [K] & $10\,000 - 13\,500$ [250] & $11\,000 - 11\,500$ & $11\,500 - 12\,000$ [100] \\
$M_*$ [$M_{\sun}$] & $0.35 - 0.80$ [0.5] & $0.40 - 0.50$ & $0.67 - 0.80$ [0.1] \\
-log$(M_\mathrm{{env}}/M_*)$ & $1.5 - 1.9$ [0.1] & $1.5 - 1.9$ & $1.5 - 1.9$ [0.1] \\
-log$(M_{\mathrm{He}}/M_*)$ & $2$ [fixed] & $2$ & $2$ [fixed] \\
-log$(M_\mathrm{H}/M_*)$ & $4 - 9$ [$1.0$] & $4 - 9$ & $4 - 9$ [$0.5$] \\ 
$X_\mathrm{{He}}$ & $0.5 - 0.9$ [0.1] & $0.5 - 0.9$ & $0.5 - 0.9$ [0.1] \\
$X_\mathrm{O}$ & $0.5 - 0.9$ [0.1] & $0.5 - 0.9$ & $0.5 - 0.9$ [0.1] \\
\hline
\end{tabular}
\end{table*}

\subsection{Results on PM~J22299+3024}
\label{sect:asteropmj}

Investigating the master grid, the best-fitting (lowest $\sigma_\mathrm{{rms}}$) model was found to be at $T_{\mathrm{eff}} = 11\,250\,$K and $M_*=0.45\,M_{\sun}$, assuming, that at least half of the modes are $\ell=1$, taking into account the better visibility of $\ell = 1$ modes over $\ell = 2$ ones (see e.g. \citealt{2008MNRAS.385..430C} and references therein). The effective temperature and mass of PM~J22299+3024 derived by optical spectroscopy is $T_{\mathrm{eff}} = 10\,630 \pm 155$\,K and $M_*=0.46 \pm 0.03\,M_{\sun}$ (log\,$g = 7.72 \pm 0.05$), respectively \citep{2015ApJS..219...19L}. That is, our model solution is hotter than we expect from optical spectroscopy, while they are in very good agreement regarding the stellar mass.

As a next step, we built a refined model grid in effective temperature, stellar mass, and the mass of the hydrogen layer, covering the parameter space in $T_{\mathrm{eff}}$ and $M_*$ around the best-fitting model found by the master grid. Table~\ref{tabl:grid} lists the parameter space we investigated by this refined grid (third column), and the corresponding step sizes (fourth column, in parentheses).

According to this refined grid, the best-fitting model has $T_{\mathrm{eff}} = 11\,400\,$K and $M_*=0.46\,M_{\sun}$. We again assumed at least three $\ell=1$ solutions for the six observed modes. In sum, this model fitting gives $\approx 800\,$K higher effective temperature than it was derived by spectroscopy, but according to the refined grid, the mass of the star found to be exactly the same as the spectroscopic solution.

\begin{figure*}
\centering
\includegraphics[width=0.47\textwidth]{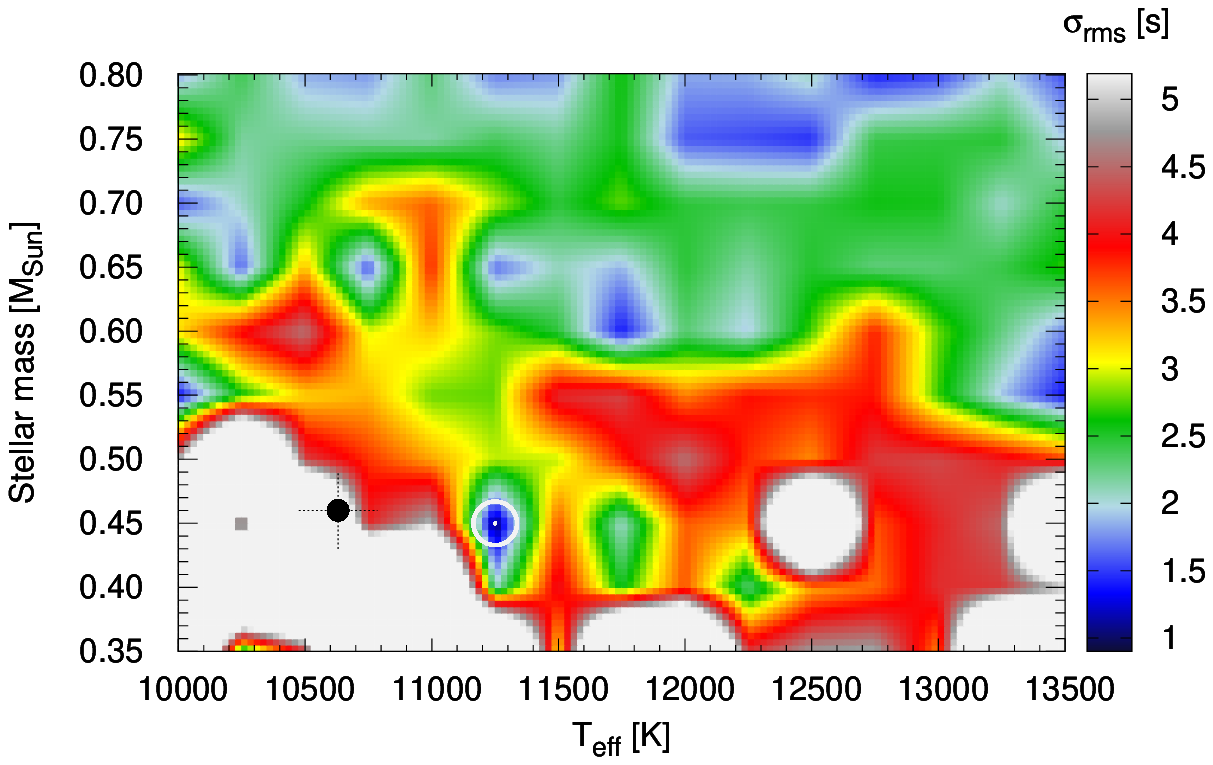}
\includegraphics[width=0.47\textwidth]{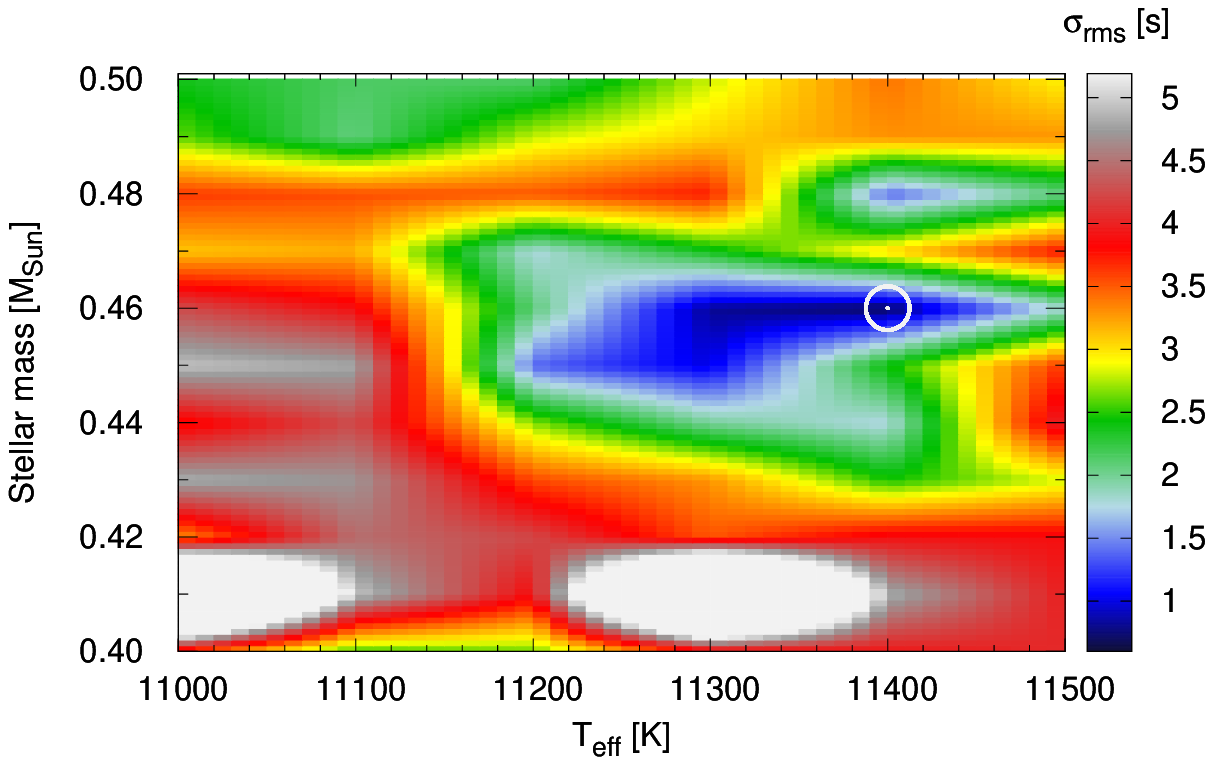}
\caption{PM~J22299+3024: models on the $T_{\mathrm{eff}} - M_*$ plane utilising the master grid (left panel) and the refined grid (right panel), assuming that at least half of the modes are $\ell = 1$. The $\sigma_\mathrm{{rms}}$ values are colour coded. The spectroscopic value is signed with a black dot, while the models with the lowest $\sigma_\mathrm{{rms}}$ values are denoted with white open circles.}{\label{fig:pmjmodels1}}
\end{figure*}

The left panel of Fig.~\ref{fig:pmjmodels1} shows the models of the master grid on the $T_{\mathrm{eff}} - M_*$ plane, assuming that at least half of the modes are $\ell = 1$. The $\sigma_\mathrm{{rms}}$ values of the period fits are colour coded. The fitting results of the refined grid are also shown in Fig.~\ref{fig:pmjmodels1} (right panel). We list the physical parameters of the two best-fitting model solutions both utilising the master and the refined grids in the first two rows of Table~\ref{tabl:modelpmj}, while Table~\ref{tabl:calcperpmj}
summarises the observed periods and the calculated periods of the best-fitting model of the refined grid.

\begin{table*}
\centering
\caption{\pmj: physical parameters of the best-fitting models.}
\label{tabl:modelpmj}
\begin{tabular}{ccccccccl}
\hline
\hline
$T_{\mathrm{eff}}$ [K] & $M_*$ [$M_{\sun}$] & -log$M_\mathrm{{env}}$ & -log$M_\mathrm{He}$ & -log$M_\mathrm{H}$ & $X_\mathrm{{He}}$ & $X_\mathrm{O}$ & $\sigma_\mathrm{{rms}}$ (s) & Comments\\
\hline
11\,250 & 0.45 & 1.7 & 2.0 & 4.0 & 0.8 & 0.8 & 0.94 & master grid\\
11\,400 & 0.46 & 1.7 & 2.0 & 4.0 & 0.8 & 0.9 & 0.67 & refined grid\\
10\,200 & 0.54 & 1.8 & 2.0 & 4.0 & 0.9 & 0.8 & 1.06 & closer to spectroscopy\\
\multicolumn{9}{l}{Spectroscopy:}\\
10\,630 & 0.46 \\
\hline
\end{tabular}
\end{table*}

\begin{table*}
\centering
\caption{PM~J22299+3024: calculated periods of the best-fitting model derived from the refined model grid.}
\label{tabl:calcperpmj}
\begin{tabular}{ccllllll}
\hline
\hline
$T_{\mathrm{eff}}$ [K] & $M_*$ [$M_{\sun}$] & \multicolumn{6}{c}{Periods in seconds ($\ell$)} \\
\hline
\multicolumn{8}{l}{Model:}\\
11\,400 & 0.46 & 1041.2 (1) & 1083.5 (1) & 1172.8 (1) & 1335.1 (1) & 1129.3 (2) & 1192.0 (2) \\
\multicolumn{8}{l}{Observations:}\\
10\,630 & 0.46 & 1041.1 & 1084.6 & 1172.8 & 1334.6 & 1130.0 & 1191.0 \\
\hline
\end{tabular}
\end{table*}

\subsection{Results on \lp}

We have a five- and a seven-period solution for the observations of \lp. Since we fit six grid parameters, we investigated the seven-period solution, including the \textit{TESS} frequencies.

Applying the master grid and assuming that at least four out of the seven modes are $\ell=1$, the best-fitting model gives $T_{\mathrm{eff}} = 11\,750\,$K and $M_* = 0.75\,M_{\sun}$, with $\sigma_\mathrm{{rms}} = 1.29\,$s. The spectroscopic values are $T_{\mathrm{eff}} = 11\,290 \pm 169$\,K and $M_*=0.65 \pm 0.03\,M_{\sun}$ (log\,$g = 8.09 \pm 0.05$), respectively \citep{2015ApJS..219...19L}. That is, we obtain a hotter and higher-mass solution. With further investigations by a refined grid around this solution, the best-fitting model is found to be at $T_{\mathrm{eff}} = 11\,900\,$K, $M_* = 0.70\,M_{\sun}$, with $\sigma_\mathrm{{rms}} = 0.75\,$s. That is, as in the case of \pmj, the asteroseismic fittings of \lp\ suggest a star hotter  than we expect from spectroscopy, but with a stellar mass not far from the spectroscopic value. The first two rows of Table~\ref{tabl:modellp} summarise the physical parameters of these best-fitting models, while Table~\ref{tabl:calcperlp} lists the calculated and observed periods of the best-fitting model of the finer grid.

\begin{figure*}
\centering
\includegraphics[width=0.47\textwidth]{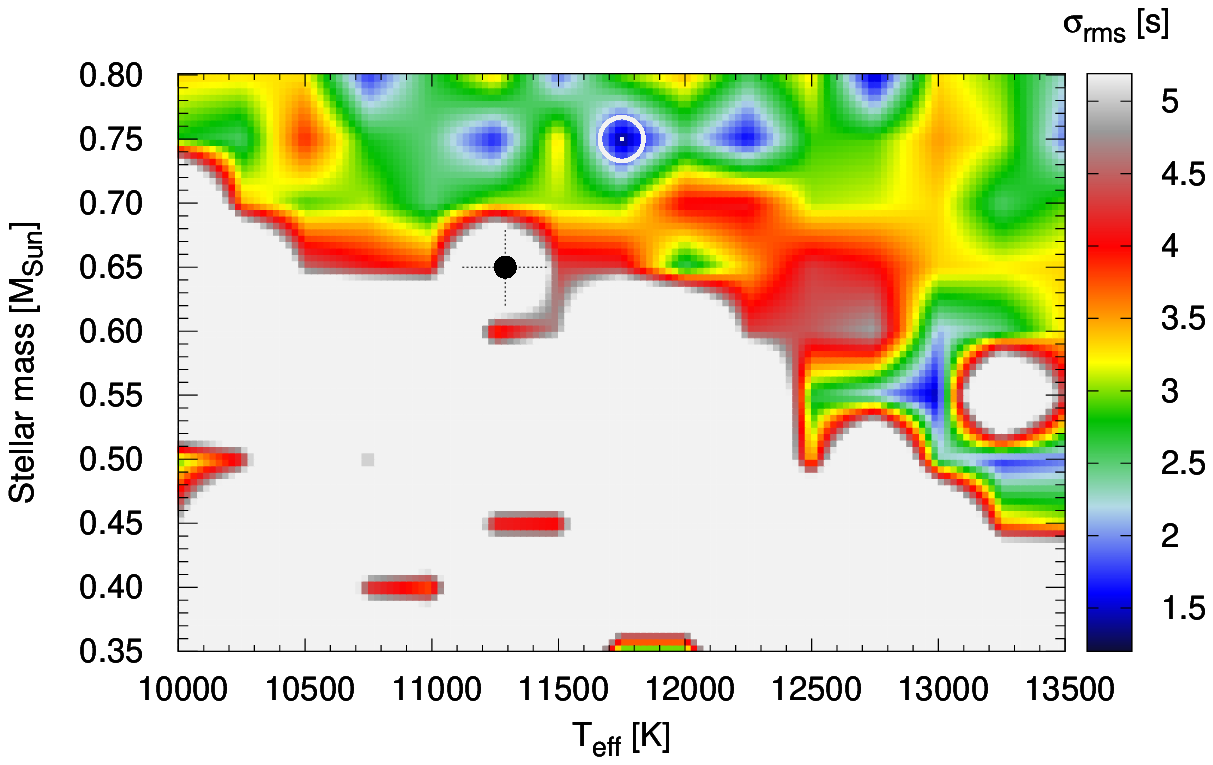}
\includegraphics[width=0.47\textwidth]{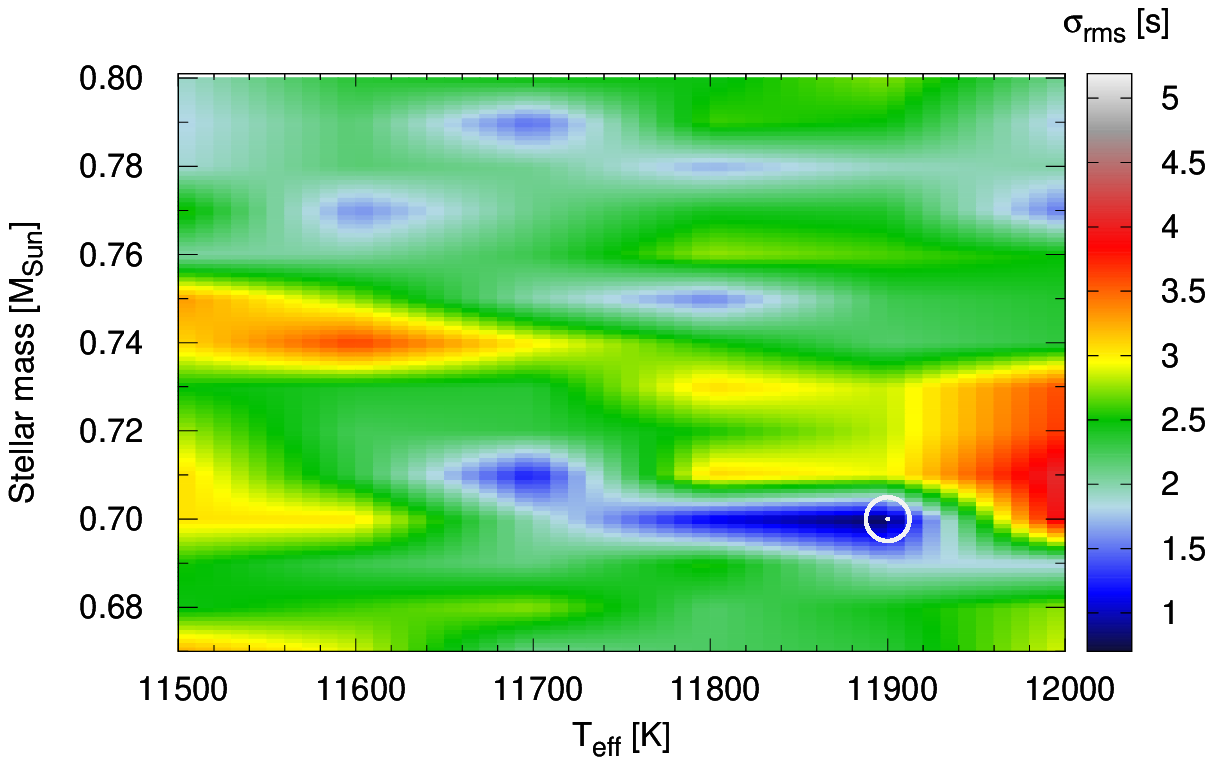}
\caption{\lp: models on the $T_{\mathrm{eff}} - M_*$ plane utilising the master grid (left panel) and the refined grid (right panel), assuming that at least four of the modes are $\ell = 1$. The $\sigma_\mathrm{{rms}}$ values are colour coded. The spectroscopic value is signed with a black dot, while the models with the lowest $\sigma_\mathrm{{rms}}$ values are denoted with white open circles.}{\label{fig:lpmodels1}}
\end{figure*}

Similarly to \pmj, we plotted the fitting results both utilising the master and the refined grids in the left and right panels of Fig.~\ref{fig:lpmodels1}, respectively.

\begin{figure*}
\centering
\includegraphics[width=0.47\textwidth]{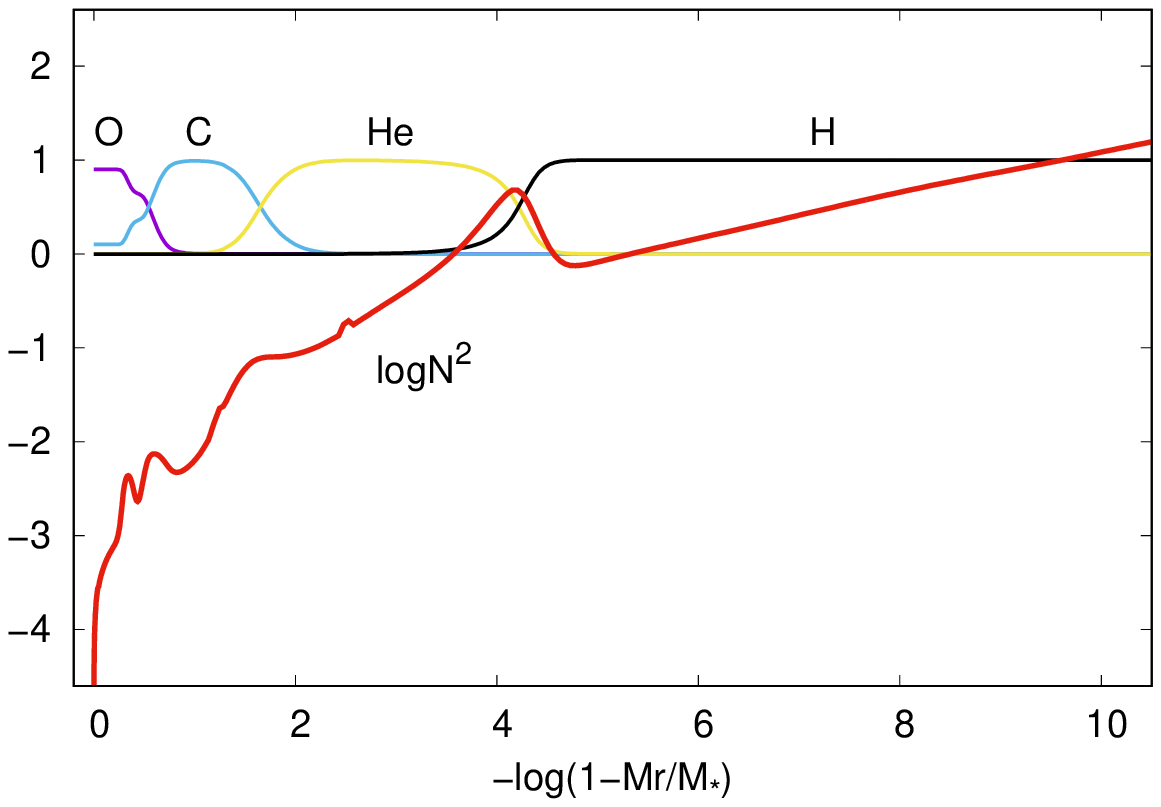}
\includegraphics[width=0.47\textwidth]{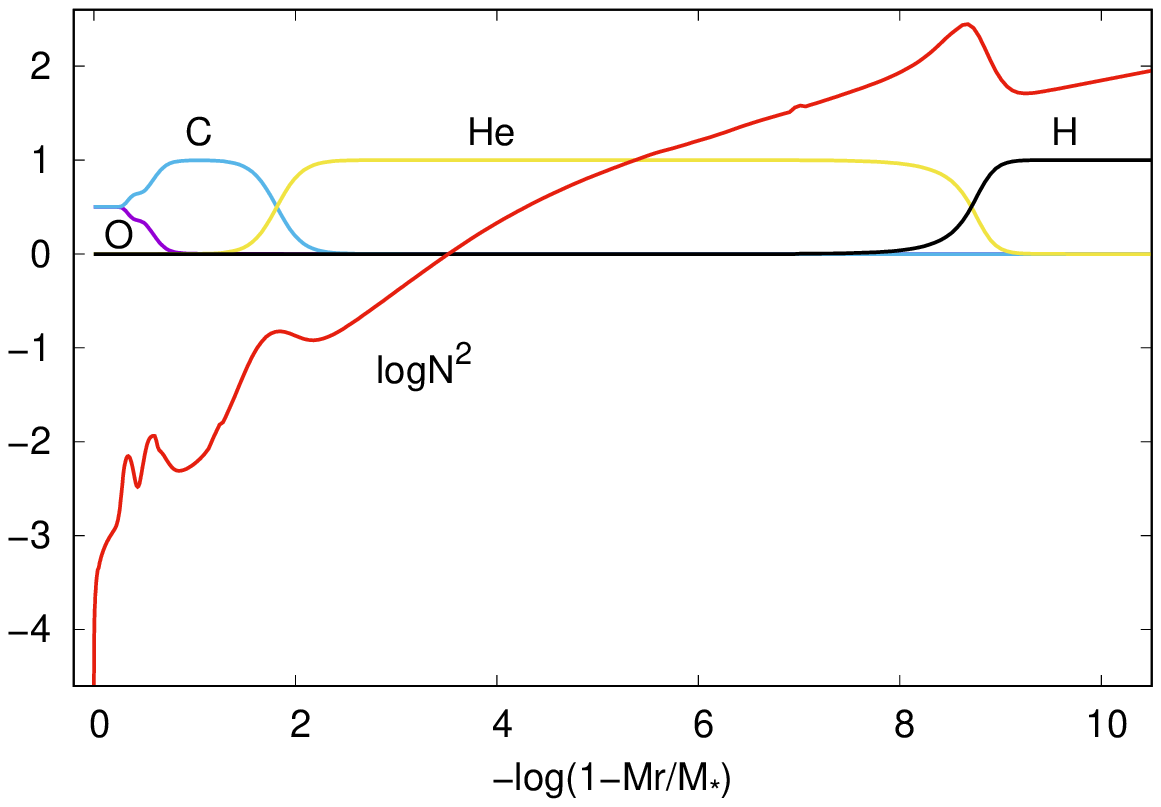}
\caption{Chemical composition profiles (in fractional abundances), and the corresponding Brunt-V\"ais\"al\"a frequencies ($\mathrm{log}\,N^2$) for the best-fitting models in the case of \pmj\ (left panel) and \lp\ (right panel), respectively. 
The model parameters for \pmj: $T_{\mathrm{eff}} = 11\,400\,$K, $M_* = 0.46\,M_{\sun}$, -log$(M_\mathrm{{env}}/M_*) = 1.7$, -log($M_\mathrm{He}/M_*)=2$, -log$(M_\mathrm{H}/M_*)=4$, $X_\mathrm{{He}}=0.8$, $X_\mathrm{O}=0.9$.
The model parameters for \lp: $T_{\mathrm{eff}} = 11\,900\,$K, $M_* = 0.70\,M_{\sun}$, -log$(M_\mathrm{{env}}/M_*) = 1.9$, -log($M_\mathrm{He}/M_*)=2$, -log$(M_\mathrm{H}/M_*)=8.5$, $X_\mathrm{{He}}=0.9$, $X_\mathrm{O}=0.5$.
}{\label{fig:chem}}
\end{figure*}

We also plotted the chemical composition profiles and the corresponding Brunt-V\"ais\"al\"a frequencies for the best-fitting models for both stars in the panels of Fig~\ref{fig:chem}. 

\begin{table*}
\centering
\caption{\lp: physical parameters of the best-fitting models.}
\label{tabl:modellp}
\begin{tabular}{ccccccccl}
\hline
\hline
$T_{\mathrm{eff}}$ [K] & $M_*$ [$M_{\sun}$] & -log$M_\mathrm{{env}}$ & -log$M_\mathrm{He}$ & -log$M_\mathrm{H}$ & $X_\mathrm{{He}}$ & $X_\mathrm{O}$ & $\sigma_\mathrm{{rms}}$ (s) & Comments\\
\hline
11\,750 & 0.75 & 1.6 & 2.0 & 4.0 & 0.9 & 0.5 & 1.29 & master grid\\
11\,900 & 0.70 & 1.9 & 2.0 & 8.5 & 0.9 & 0.5 & 0.75 & refined grid\\
11\,800 & 0.70 & 1.9 & 2.0 & 8.5 & 0.5 & 0.6 & 1.07 & closer to spectroscopy\\
\multicolumn{9}{l}{Spectroscopy:}\\
11\,290 & 0.65 \\
\hline
\end{tabular}
\end{table*}

\begin{table*}
\centering
\caption{\lp: calculated periods of the best-fitting model derived from the refined model grid.}
\label{tabl:calcperlp}
\begin{tabular}{cclllllll}
\hline
\hline
$T_{\mathrm{eff}}$ [K] & $M_*$ [$M_{\sun}$] & \multicolumn{6}{c}{Periods in seconds ($\ell$)} \\
\hline
\multicolumn{8}{l}{Model:}\\
11\,900 & 0.70 & 767.8 (1) & 820.0 (1) & 890.9 (1) & 979.2 (1) & 739.4 (2) & 849.1 (2) & 900.3 (2) \\
\multicolumn{8}{l}{Observations:}\\
11\,290 & 0.65 & 767.5 & 820.4 & 890.1 & 978.4 & 739.3 & 847.6 & 900.2 \\
\hline
\end{tabular}
\end{table*}

\subsection{Models closer in effective temperature and stellar mass to the spectroscopic solutions}

For comparison, we covered the $T_{\mathrm{eff}}$ -- $M_*$ parameter space in the $\pm 3\sigma$ vicinity of the spectroscopic solutions both for \pmj\ and \lp. That is, this grid for \pmj\ covers the parameter range of $10\,200-11\,100$\,K in effective temperature, and $0.37-0.55\,M_{\sun}$ in stellar mass. The step sizes were the same as we used for the refined grids described before. In the case of \lp, we covered the effective temperature and stellar mass parameter ranges of $10\,800 - 11\,800$\,K and $0.56 - 0.74\,M_{\sun}$, respectively.

In the case of \pmj, assuming that at least half of the modes are $\ell = 1$, the model with the lowest $\sigma_\mathrm{{rms}}$ has $T_{\mathrm{eff}} = 10\,200$\,K, and $M_* = 0.54\,M_{\sun}$ ($\sigma_\mathrm{{rms}} = 1.06$\,s). 

For \lp, the best-fitting model utilising seven modes with at least four $\ell=1$ ones, has $T_{\mathrm{eff}} = 11\,800\,$K, and $M_* = 0.70\,M_{\sun}$ ($\sigma_\mathrm{{rms}} = 1.07\,$s)

For completeness, we also list the physical parameters of these two model solutions in Tables~\ref{tabl:modelpmj} and \ref{tabl:modellp}. We can see that the lowest $\sigma_\mathrm{{rms}}$ values belong to the models we found utilising the refined grid but not taking into account the spectroscopic solutions. 

\subsection{Asteroseismic distances}

There is an excellent way to validate our asteroseismic solutions: by comparing the seismic distances calculated by the models with the astrometric distances provided by the \textit{Gaia} space mission \citep{2016A&A...595A...1G}, see e.g. the example in \citet{2019A&A...632A..42B}. In calculating a seismic distance, at first we have to check the luminosity of the model. Knowing the luminosity value $\mathrm{log}\,(L/L_{\sun})$, and the bolometric magnitude of the Sun ($M_{bol,\sun} = 4.74$), we can derive the bolometric magnitude of the star using the correlation $M_{bol} = M_{bol,\sun} - 2.5 \mathrm{log}(L/L_{\sun})$. Now we need the bolometric correction factor (BC) to calculate the absolute visual magnitude of the star: $M_V = M_{bol} - \mathrm{BC}$. \citet{1995PASP..107.1047B} performed colour-index and magnitude calculations using DA and DB model grids. According to table 1 in \citet{1995PASP..107.1047B}, $\mathrm{BC} = -0.441$ and $-0.611$ at temperatures $11\,000$ and $12\,000$\,K, respectively. From this, we derived the bolometric corrections to the actual temperatures with linear interpolations. Next we need the apparent visual magnitude ($m_V$) of the star to apply the distance modulus formula and derive the seismic distance with the given model parameters. Following \citet{2019A&A...632A..42B}, we utilised the Fourth US Naval Observatory CCD Astrograph Catalog \citep{2012yCat.1322....0Z} to find the apparent visual magnitude of the star. At last, we compared the seismic distance derived this way with the \textit{Gaia} early third release (\citealt{2020arXiv201201533G}, hereafter EDR3) geometric distance value published by \citet{2021AJ....161..147B}.

Table~\ref{tabl:dist} summarises the results of the different steps in deriving the seismic distances both for \pmj\ and \lp, respectively. We found for \pmj, that the $T_{\mathrm{eff}} = 11\,400\,$K and $M_*=0.46\,M_{\sun}$ model of the refined grid provides a seismic distance equal within the errors with the \textit{Gaia} astrometric distance, that is, they are in excellent agreement. The seismic distance of the much cooler model with $T_{\mathrm{eff}}$ and $M_*$ within the $3\sigma$ vicinity of the spectroscopic values suggests a star about 20\,pc closer to us. For \lp, both model solutions provide seismic distances in good agreement with the \textit{Gaia} distance. Note that they have similar physical parameters, thus in this case, we can conclude that both model solutions are acceptable considering the seismic and astrometric distances.  

\begin{table*}
\centering
\caption{Steps in deriving the seismic distances of the stars. We listed the parameters for two models in the case of both stars: the first models belong to the best-fitting models not considering the spectroscopic solutions, while the second models belong to models found in the $\pm3\sigma$ vicinity of the spectroscopic $T_{\mathrm{eff}}$ and $M_*$ values.
We also list the \textit{Gaia} EDR3 geometric distance values for comparison.}
\label{tabl:dist}
\begin{tabular}{rrrrr}
\hline
\hline
 & \multicolumn{2}{c}{\pmj} & \multicolumn{2}{c}{\lp} \\
\hline
$T_{\mathrm{eff}}$ [K] & $11\,400$ & $10\,200$ & $11\,900$ & $11\,800$\\
$M_*$ [$M_{\sun}$] & 0.46 & 0.54 & 0.70 & 0.70\\
$\mathrm{log}L/L_{\sun}$ & $-2.453$ & $-2.742$ & $-2.652$ & $-2.667$\\
$M_{bol}$ [mag] & $10.873$ & $11.595$ & $11.370$ & $11.408$\\
BC [mag] & $-0.509$ & $-0.352$ & $-0.594$ & $-0.577$\\
$M_V$ [mag] & $11.382$ & $11.947$ & $11.964$ & $11.985$\\
$m_V$ [mag] & \multicolumn{2}{c}{$16.161\pm0.01$} & \multicolumn{2}{c}{$15.294\pm0.09$} \\
$d_{seismic}$ [pc] & $90.34\pm0.42$ & $69.62\pm0.32$ & $46.3\pm1.9$ & $45.9\pm1.9$\\
$d_{Gaia}$ [pc] & \multicolumn{2}{c}{$90.92^{+0.55}_{-0.40}$} & \multicolumn{2}{c}{$45.97\pm0.09$} \\
\hline
\end{tabular}
\end{table*}

\section{Summary and discussion}

In this paper, we present the pulsational characteristic of two ZZ~Ceti stars; our newly discovered \pmj, and the already known \lp. Both stars show complex pulsational behaviour, revealed by the Fourier transforms of their data sets with several possible pulsation peaks. 

We found that both \pmj, and \lp\ show several pulsational frequencies not known before. With more ground-based observations or measurements from the space, they are good candidates to become pulsators rich in known frequencies, that is, with a dozen or more discovered pulsation modes. Considering the usual amplitude and phase variations observed in these type of stars, our observations represents an important snapshot of the current pulsational behaviour of these targets.

In the case of \pmj, we accepted six pulsation modes in the $1041 - 1335\,$s period range, and further Fourier amplitude peaks were identified as possible pulsation frequencies. Similarly, in the case of \lp, five pulsation modes were accepted in the $768 - 978\,$s period range, but also further modes may be present in the ground-based data set. The \textit{TESS} space telescope also observed this star, and we were able to complement the set of accepted pulsation modes with two additional ones, while the identifications of a third \textit{TESS} pulsation component confirmed the ground-based detection of the same mode.

There were two important difficulties when we derived the pulsation frequencies.
The first one was the usage of single-site ground-based observations, where the 1\,d$^{-1}$ aliases appear with relatively large amplitudes in the Fourier transforms. This made the identifications of the pulsation peaks ambiguous, especially because the alias structures of closely situated pulsation peaks overlapped, both for \pmj\ and \lp. Another source of the ambiguities was the amplitude and phase variations of the pulsation modes over time-scales shorter than the duration of the observations. This can lead to the emergence of additional peaks, appearing as extended line widths in the Fourier transforms of the data sets. This effect was clearly demonstrated e.g. in the case of the \textit{Kepler} observations of ZZ~Ceti stars showing pulsation modes longer than $\sim 800\,$s (which is the case both for \pmj\ and \lp), presented by \citet{2017ApJS..232...23H}.

Beyond the frequency analyses of the data sets on these two stars, we performed asteroseismic investigations of both objects. For \pmj, our best model solution has an effective temperature and stellar mass of $11\,400\,$K and $0.46\,M_{\sun}$, respectively. The stellar mass is the same as the value provided by spectroscopy, but this model is almost $800\,$K hotter than the spectroscopic solution. However, the seismic distance calculated for our best-fitting model is in excellent agreement with the astrometric distance derived by \textit{Gaia} observations, supporting our results. 

Note that in our asteroseismic analysis we calculated periods of model white dwarfs assuming carbon and oxygen (C/O) core. However, in the case of \pmj, both its asteroseismic and spectroscopic mass fall just between the mass ranges of the helium-core and C/O-core white dwarfs. The so-called low-mass ($M_* \leq 0.45\,M_{\sun}$) white dwarf stars are expected to have helium cores and being results of evolution in binary systems, see e.g. \citet{2016MNRAS.455.3413K} and references therein, or the evolutionary calculations focusing on helium-core white dwarfs by \citet{2013A&A...557A..19A}. 

The model fits of \lp\ give effective temperatures in the range of $11\,800 - 11\,900\,$K, that is, \lp\ is more likely to be around the middle of the ZZ~Ceti instability strip, rather than close to the red edge, as was suggested by the spectroscopic observations. Considering its mass, we find solutions with $0.70\,M_{\sun}$, which is near to the $0.65\pm0.03\,M_{\sun}$ spectroscopic value. Similarly to the case of \pmj, the seismic and astrometric distances are in good agreement.

We compared the central abundances of carbon and oxygen of our best-fitting models on \lp\ with the predictions on these parameters based on stellar evolutionary calculations published by \citet{2012MNRAS.420.1462R}. We utilised their data base\footnote{http://evolgroup.fcaglp.unlp.edu.ar/TRACKS/PULSATIONS/PULSATIONS\_DA/pulsations\_cocore\_par.html}, and found a stellar model with physical parameters close to our best-matching solutions: it has $T_{\mathrm{eff}} = 11\,814\,$K, $M_*=0.705\,M_{\sun}$, $-\mathrm{log}(M_{\mathrm H}/M_*)= 8.34$, and central oxygen abundance of $X_{\mathrm O} = 0.66$. 
That is, evolutionary predictions prefer higher central oxygen abundance than we obtained for our two best-fitting models. Our best model near the spectroscopic parameters, which gives a core oxygen content of $X_{\mathrm O} = 0.6$ more in accordance with evolutionary predictions, has only slightly higher $\sigma_\mathrm{rms} = 1.07$\,s compared to the very best-fitting one without spectroscopic restriction, which has a lower core oxygen abundance of $X_{\mathrm O} = 0.5$ and $\sigma_\mathrm{rms} = 0.75$\,s. We cannot distinguish between the two best models based on their asteroseismic distances, since they agree with one another and with the {\it Gaia} distance very well.


\begin{acknowledgements}
The authors thank the anonymous referee for the constructive comments and recommendations on the manuscript.

The authors acknowledge the financial support the Lend\"ulet Program of the Hungarian Academy of Sciences, projects No. LP2018-7/2020 and LP2012-31.

ZsB acknowledges the support provided from the National Research, Development and Innovation Fund of Hungary, financed under the PD$_{17}$ funding scheme, project no. PD-123910, and the support by the J\'anos Bolyai Research Scholarship of the Hungarian Academy of Sciences.

CsK acknowledges the support provided from the \'UNKP-20-2 New National Excellence Program of the Ministry of Human Capacities.

This paper includes data collected with the TESS mission, obtained from the MAST data archive at the Space Telescope Science Institute (STScI). Funding for the TESS mission is provided by the NASA Explorer Program. STScI is operated by the Association of Universities for Research in Astronomy, Inc., under NASA contract NAS 5–26555.

This work has made use of data from the European Space Agency (ESA) mission {\it Gaia} (\url{https://www.cosmos.esa.int/gaia}), processed by the {\it Gaia} Data Processing and Analysis Consortium (DPAC, \url{https://www.cosmos.esa.int/web/gaia/dpac/consortium}). Funding for the DPAC
has been provided by national institutions, in particular the institutions participating in the {\it Gaia} Multilateral Agreement.
\end{acknowledgements}



\bibliographystyle{aa} 
\bibliography{exploring} 

\end{document}